\documentclass{article}
\pdfoutput=1

\PassOptionsToPackage{numbers}{natbib}
\usepackage{titletoc}
\usepackage{hyperref}
\usepackage[page,header]{appendix}
\usepackage[final]{nips_2017}
\usepackage[utf8]{inputenc} 
\usepackage[T1]{fontenc}    
\usepackage{url}            
\usepackage{booktabs}       
\usepackage{amsfonts}       
\usepackage{nicefrac}       
\usepackage{microtype}      
\usepackage{xcolor}
\usepackage{hyperref}
\usepackage{graphicx}
\usepackage{algorithmicx}
\usepackage{algorithm}
\usepackage{algpseudocode}
\usepackage{mathrsfs}
\usepackage{amssymb}
\usepackage{graphicx,wrapfig,lipsum}
\usepackage{natbib}
\usepackage{caption}
\usepackage{float}
\usepackage{subcaption}
\usepackage{enumitem}
\usepackage{siunitx}
\usepackage{multirow}
\usepackage{tikz}
\usepackage{footnote}
\usetikzlibrary{arrows,shapes,snakes,automata,backgrounds,petri,positioning,calc}
\usepackage{lipsum}
\usepackage[multiple]{footmisc}
\usepackage{placeins}

\newif\ifshort
\shortfalse

\ifshort
    \usepackage{titlesec}
    \titlespacing\section{0pt}{0pt plus 4pt minus 2pt}{0pt plus 2pt minus 2pt}
    \titlespacing\subsection{0pt}{0pt plus 4pt minus 2pt}{0pt plus 2pt minus 2pt}
\else
    
\fi

\title{ORRB - OpenAI Remote Rendering Backend}

\author{
  Maciek Chociej~~~~Peter Welinder~~~~Lilian Weng \\
  OpenAI
}

\begin{document}
\maketitle

\begin{figure}[h!]
\centering
\includegraphics[width=1.0\textwidth]{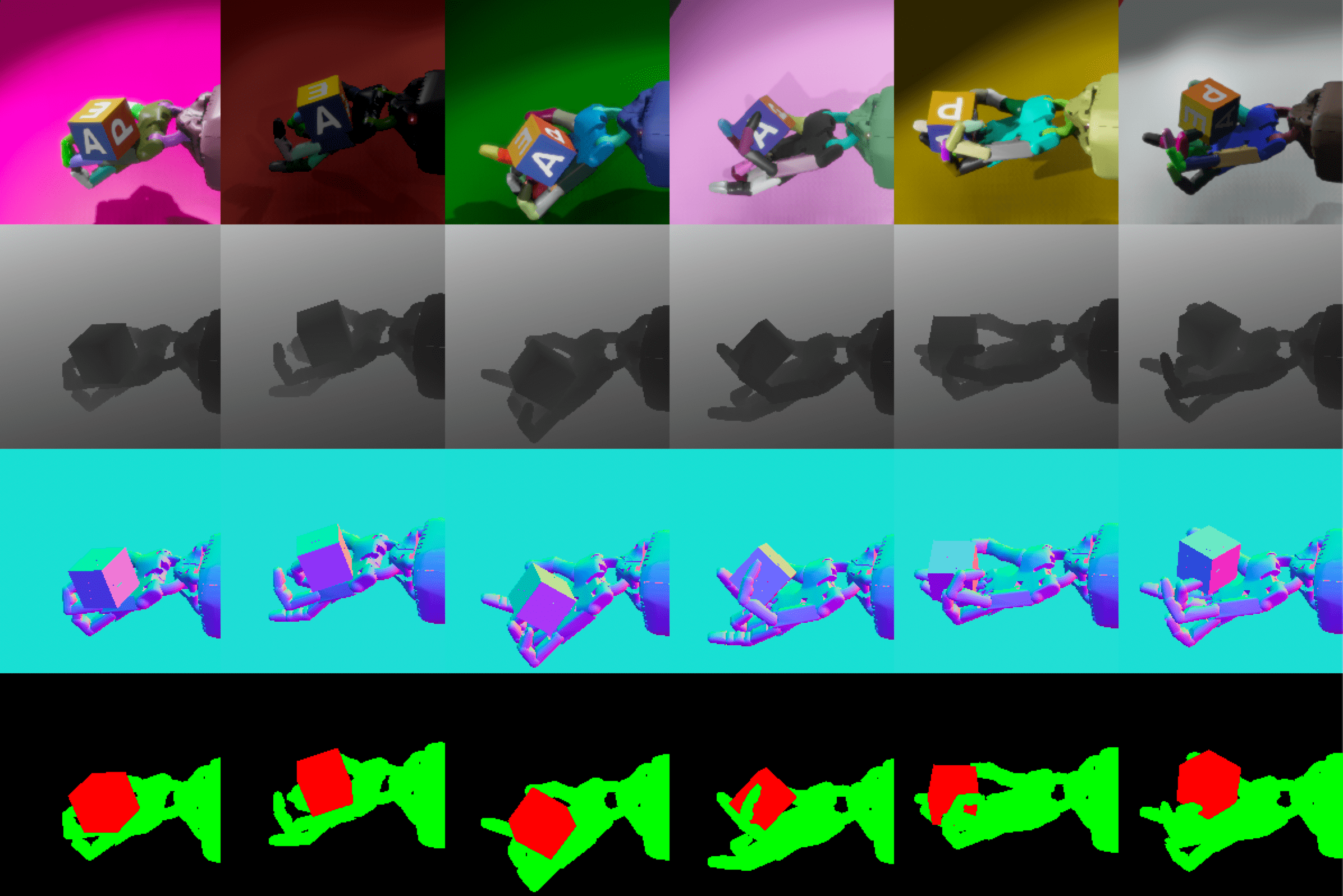} \\ \vspace{0.3cm}
\caption{A batch of visually randomized samples (RGB, depth, normal, and segmentation channels), rendered with the OpenAI Remote Rendering Backend (ORRB), depicting the Shadow Dexterous Hand manipulating a colourful block.}
\end{figure}

\begin{abstract}
We present the OpenAI Remote Rendering Backend (ORRB), a system that allows fast and customizable rendering of robotics environments. It is based on the Unity3d \cite{unity} game engine and interfaces with the MuJoCo \cite{mujoco} physics simulation library. ORRB was designed with visual domain randomization in mind. It is optimized for cloud deployment and high throughput operation. We are releasing it to the public under a liberal MIT license: \url{https://github.com/openai/orrb}.
\end{abstract}

\startcontents[mainsections]

\section{Introduction}
\label{sec:introduction}
Simulation-to-reality (Sim2Real) transfer is a central problem in applying reinforcement learning (RL) to real world robotics tasks. Many complex problems that can be solved in simulation fail to transfer to physical robots due to unmodeled effects or uncalibrated parameters. Systems that depend on computer vision also suffer from this fate. It often hard to transfer from synthetic data to a controlled lab setup, not to mention handling the endless variety of lighting conditions, material appearance and optical phenomena in a robust way.

Domain randomization \cite{tobin2017domain} \cite{DBLP:conf/rss/SadeghiL17} is a promising approach to closing the reality gap. Unlike striving for photo-realism or using real data in the training pipeline, it is simple, fast and can be used to produce data in abundant amounts. However, there are few good existing tools to perform visual domain randomization.

We have created ORRB with the following goals in mind:

\begin{itemize}
    \item \textbf{Domain Randomization as First-Class Citizen:} expose appearance control parameters externally and allow randomization on different level of granularity, 
    \item \textbf{Modular, Customizable \& Expandable Architecture:} provide a framework that enables rapid prototyping and development of new randomization and augmentation techniques,
    \item \textbf{Performance:} optimize code for large scale, high throughput, batch rendering in a distributed, headless environment,
    \item \textbf{Ease of Use:} allow seamless, easy to configure integration with Machine Learning (ML) libraries.
\end{itemize}

We have used it to successfully train Dactyl \cite{DBLP:journals/corr/abs-1808-00177} -- the dexterous robotic hand -- purely from simulated, synthetic data.

The paper is structured as follows. \autoref{sec:related_work} gives an overview of previous work in the field, especially the use of game engines to generate synthetic training data. \autoref{sec:characteristics} elaborates on the characteristics of the typical robotics RL setup we work with, and how that influences the rendering pipeline. \autoref{sec:technical_details} focuses on system design and technical details of implementation. Finally,  \autoref{sec:performance} contains a performance evaluation across different setups.

\section{Related Work}
\label{sec:related_work}
Computer games are a natural choice for machine learning research. They present rich, interactive environments, with simple, general mechanics that often give rise to complex emergent gameplay. The skills needed to master different games range from impulse reflexes required to solve an arcade game, to long term planning and strategic reasoning that is necessary to compete on a professional level in complex real-time strategy titles. This is why recent results like: Atari \cite{DBLP:journals/corr/MnihKSGAWR13}, Dota 2 \cite{dota2} or Starcraft II \cite{alphastarblog} were met with great interest, and mark the steady progress of ML/RL techniques.

A second track of research involves using game engines and middleware to craft environments that can be used to solve non-game related problems. The most prominent application is autonomous navigation. Both Unity \cite{unity} and Unreal \cite{unreal} game engines are used in autonomous drone \cite{DBLP:journals/corr/ShahDLK17} and car \cite{simviz} simulations.

Finally custom designed environments built on top of games and game engines like: Doom \cite{DBLP:journals/corr/KempkaWRTJ16}, Quake III Arena \cite{Jaderberg859}, Minecraft \cite{DBLP:journals/corr/AbelADKS16} or Unity \cite{journals/corr/abs-1902-01378,journals/corr/abs-1809-02627} offer a testbed for basic ML research.

\section{Game vs. Robotics RL Workloads}
\label{sec:characteristics}
Modern game engines are some of the most complicated pieces of software. Not only they are on the cutting edge of real-time computer graphics and physical simulation, but also must provide flexibility to empower the creative process behind the coolest games. They need to execute on a wide variety of different hardware and operating system platforms, and support enjoyable entertainment on a spectrum of devices with computing capabilities ranging from mobile phones to multi-GPU desktop computers.

Game engines are however designed with some characteristics that make them impractical in large scale machine learning. They are optimized to generate high resolution images with resolution and frequency aimed at human perceptive capabilities. A typical game produces $30$ to $60$ images per second at a resolution in the order of $1920 \times 1080$ pixels. A game engine is outputting the images in a serial way, one image after another, interactively controlled by the human operator, keeping the temporal coherence in order to produce the illusion of smooth motion. Low latency and consistent rendering rate are most important to provide pleasant experience. The rendering system is most often GPU constrained in graphic heavy titles.

In large scale ML scenarios we rarely require low latency. During training we can often exploit the large data parallelism of training samples and trajectories. Because of this embarrassing parallelism we can benefit from batch rendering and reduce some of the per frame and communication overheads. Total rendering throughput proves to be the more important metric. For computer vision purposes we require low resolution images, common sizes include: $64 \times 64$, $112 \times 112$, $224 \times 224$. This in turn moves the compute bottleneck from GPU rasterization and shading, and towards the work necessary to set up the rendering on the CPU. 

A typical game environment involves multiple dynamic objects interacting in a highly customized ways. The description of the state can be arbitrarily complex. In a robotic simulation we usually operate in the framework of Markov Decision Processes (MDP) on a number of rigid and soft bodies connected by tendons and joints. This allows for compact state description. Furthermore we exploit the inherent parallelism of our setting and perform computation in a Single Scene Multiple States (SSMS) way. The bulk of the data, that is geometry, textures, materials, and the kinematic hierarchy description can be shared between instances and only the randomized parameters and the lightweight state need to be kept separate.

Most of the system design and optimization effort was put into working around the limitations of the interactive mode of operation in the Unity game engine and optimizing it for high throughput, batch, SSMS rendering.

\section{System Design \& Technical Details}
\label{sec:technical_details}
\subsection{Renderer as a Service}

ORRB renderer is a standalone binary built with the Unity game engine. On start-up the binary loads a scene XML and all the necessary assets, i.e: textures and meshes. Then, the application reads problem specific configuration files. Those contain the description of how the scene should be transformed and randomized for rendering. They also describe how the state description maps to the kinematic hierarchy of the reconstructed scene. Finally ORRB starts a GRPC\cite{grpc} service on a predefined port and awaits for incoming update and batch render requests.

\subsection{Scene Import and Configuration}

For scene description we use the MuJoCo XML format. We support a large subset of the specification, including: bodies; geoms both of mesh and primitive types; hinge, slide and ball joints; cameras; and sites. We load the material definitions and use the various parameters like: RGB color, transparency specular and emissive strength to set up Unity materials' appearances. We also support the property class inheritance and XML includes. The mesh assets can be loaded from both binary and text STL files.

In addition to the XML scene description we use configuration files that contain the description of the transformations and randomizations to be applied. The configuration files are stored as human readable YAML protocol buffers. We use the same protocol buffers, albeit in binary wire format, to update the renderer configuration. Configuration updates are executed on the fly, over RPC. The ability to change randomization parameters during training opens new possibilities and makes adaptive randomization and similar techniques possible.

\subsection{Component Manager}

\begin{figure}[h!]
\centering
\includegraphics[width=1.0\textwidth]{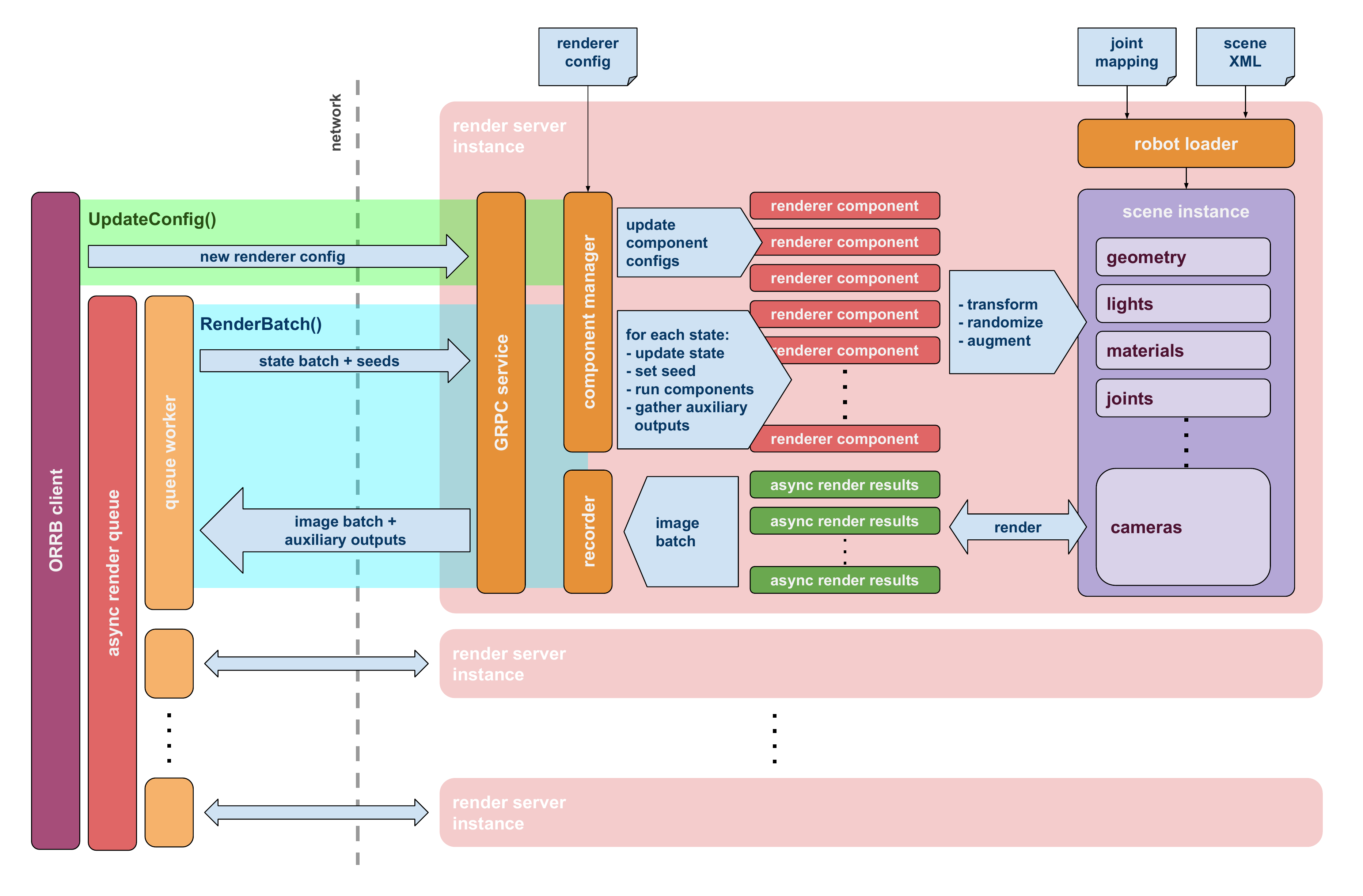} \\ \vspace{0.3cm}
\vspace{0.3cm}
\caption{System overview.}
\end{figure}

The component manager is the central part of ORRB. Its main task is instantiating, running and updating the renderer components. A renderer component is an encapsulated module responsible for single scene augmentation or randomization. We use the same framework for some interactive tools. A renderer component can be attached to any entity in the scene's kinematic hierarchy with the goal of modifying the properties of that singular entity or the entities in its kinematic sub-tree. Sub-entities can be filtered by type and name. This framework also allows the renderer components to emit auxiliary outputs, which can be used both in training auxiliary tasks and in rendering adaptation techniques.

To the extent possible, randomization and rendering is deterministic. A seed can be provided with every batch, or every state. The seed is used to reinitialize the random number generator. We put diligent care into making sure that the randomizers are executed in the same order, and that the flow of the randomized code is exactly the same -- leading to reproducible, identical randomization results. That said, in our current model, with OpenGL rendering and a number of GPU based rendering techniques, some low level pixel discrepancy of the final images is unavoidable. 

We release ORRB with a number of ready-made renderer components. Those can be roughly divided in two categories: randomizers and scene setup utility components.

\subsubsection{Randomizer Components}

\begin{itemize}
    \item \textbf{MaterialRandomizer} -- randomizes the tint, metallic and glossiness levels in materials. Additionally can assign random pattern and normal textures with randomized tiling and offset parameters,
    \item \textbf{FixedHueMaterialRandomizer} -- randomizes the material appearance near some predefined, calibrated values. This randomizer operates in HSV color space and allows specifying separate randomization radiuses and clamping ranges for the three components. Additionally metallic level, glossiness, e†missive probability and emissive power can be randomized within configurable ranges,
    \item \textbf{CameraRandomizer} -- randomizes the position, orientation and angle of view of the recording cameras. This randomizer operates in two modes: \textit{Jitter} -- where a small random perturbations are applied, and \textit{Orbit} -- where the camera is randomly placed orbiting a point of interest specified in the scene. The two modes can be both applied at the same time,
    \item \textbf{JointRandomizer} -- randomizes joint positions within their limit ranges,
    \item \textbf{LightRandomizer} -- randomizes the intensity of scene lights. Allows specifying the range from which the total scene illumination will be drawn. Exposes configurable ranges for the random relative weights of each light's individual intensity contributions, and spotlight angles.
    \item \textbf{PostprocessingRandomizer} -- randomizes the post processing effect parameters. We use the Unity PostProcessingV2 stack, and allow customized randomization ranges for: \textit{Ambient Occlusion} intensity; \textit{Color Grading} hue, saturation, contrast, brightness, temperature, and tint; \textit{Bloom} intensity, and diffusion radius; and finally \textit{Grain} intensity, size, and coloration. 
\end{itemize}

\subsubsection{Utility Components}

\begin{itemize}
    \item \textbf{Tracker} -- generates auxiliary outputs with screen space coordinates and bounding boxes for the tracked objects,
    \item \textbf{TranslateRotateScale} -- modifies the local position, rotation, and scale of an entity in the kinematic hierarchy,
    \item \textbf{Hide} -- enables and disables the rendering of a given entity,
    \item \textbf{LookAt} -- orients the entity, pointing towards a specified target,
    \item \textbf{LightSetup} -- creates a number of scene lights in a programmatic way. Allows specifying the ranges for: distance from, and height above a specified target. 
\end{itemize}

\subsubsection{Camera Calibrator}

The Camera Calibrator renderer component can be used to align the scene cameras with the footage from the physical ones. It is used in the interactive mode, and provides a rudimentary GUI to fine-tune the camera's position, rotation and angle of view.

\subsection{Capture Pipeline}

In order to achieve high throughput we removed unnecessary data stalls and GPU / CPU synchronization. The capture pipeline maintains a pool of render textures that are used in a round robin order. Similarly the destination textures for the whole batch are pre-allocated, and transferred back to the CPU memory when the whole batch is ready, with bulk DMA. Because consequent frames have no data dependencies, rendering finalization on the GPU can happen in an asynchronous, pipelined way.

We support 4 rendering modes: the main RGB image, optionally with additional transparency channel; segmentation map; depth map; and the screen space normal vector map. 

\subsection{Client}

ORRB is to large extent client language agnostic. Any language that supports GRPC can communicate with the render service. We provide simple client side utilities for Python: server life-cycle management, and parallel batch render executor. Additionally we release a small Python suite of demos, benchmarks and unit tests.

\subsection{Deployment}

We have successfully ran ORRB both on local machines operating under OSX and Ubuntu, and in server environments i.e. on Kubernetes, Microsoft Azure and Google Cloud Platform virtual machines. True headless rendering is currently not supported in Unity. In order to make datacenter rendering possible we start a Xorg instance on top of Nvidia virtual devices. Then, we set up multiple virtual screens (in \textit{xorg.conf}) to efficiently control multi-GPU rendering. 

\section{Performance}
\label{sec:performance}
In order to measure the performance of our system, we have used the environment from \cite{DBLP:journals/corr/abs-1808-00177}. The scene depicts a Shadow Dexterous Hand holding a colorful block. It contains three cameras and three spot lights with soft shadow casting enabled. The renderer configuration consists of: calibrated material randomizers for the block, a material randomizer for the hand and the background, a camera jitter randomizer, a light position randomizer, a light intensity randomizer, and finally the postprocessing effects randomizer. We use a batch size of $64$ samples, which amounts to $192$ images rendered in one call. The images are rendered in $200 \times 200$ pixels resolution in $3 \times 8$ bit RGB color depth.

Measurements have been performed on Google Cloud Platform. The benchmark machine was based on a n1-standard-96 VM with a $96$ virtual core processor and $8$ NVidia V100 GPUs. We used Ubuntu 16.04, NVidia 410.48 drivers; and Unity 2018.3 was used to build the render server binary. Benchmark results were produced with the \textit{benchmark.py} script shipped with ORRB.

The analysis of the performance benchmarks shows a number of different interesting effects. First of all, it takes between $16$ and $24$ render servers running in parallel to fully saturate a V100 GPU. As the number of render servers increases the single threaded python model becomes the bottleneck and the client code begins to throttle performance. In order to counter that, we run multiple, separate client processes --- one per approximately $8$ to $12$ render server instances. With this approach we were able to scale up to 3 GPUs. Further scaling is limited by insufficient CPU processing power.

\begin{table}[ht!]
    \centering
    \caption{
    Performance of ORRB on $1-4$ V100 GPUs with different numbers of render servers and MPI clients working in parallel.}
    \renewcommand{\arraystretch}{1.3}
    \includegraphics[width=1.0\textwidth]{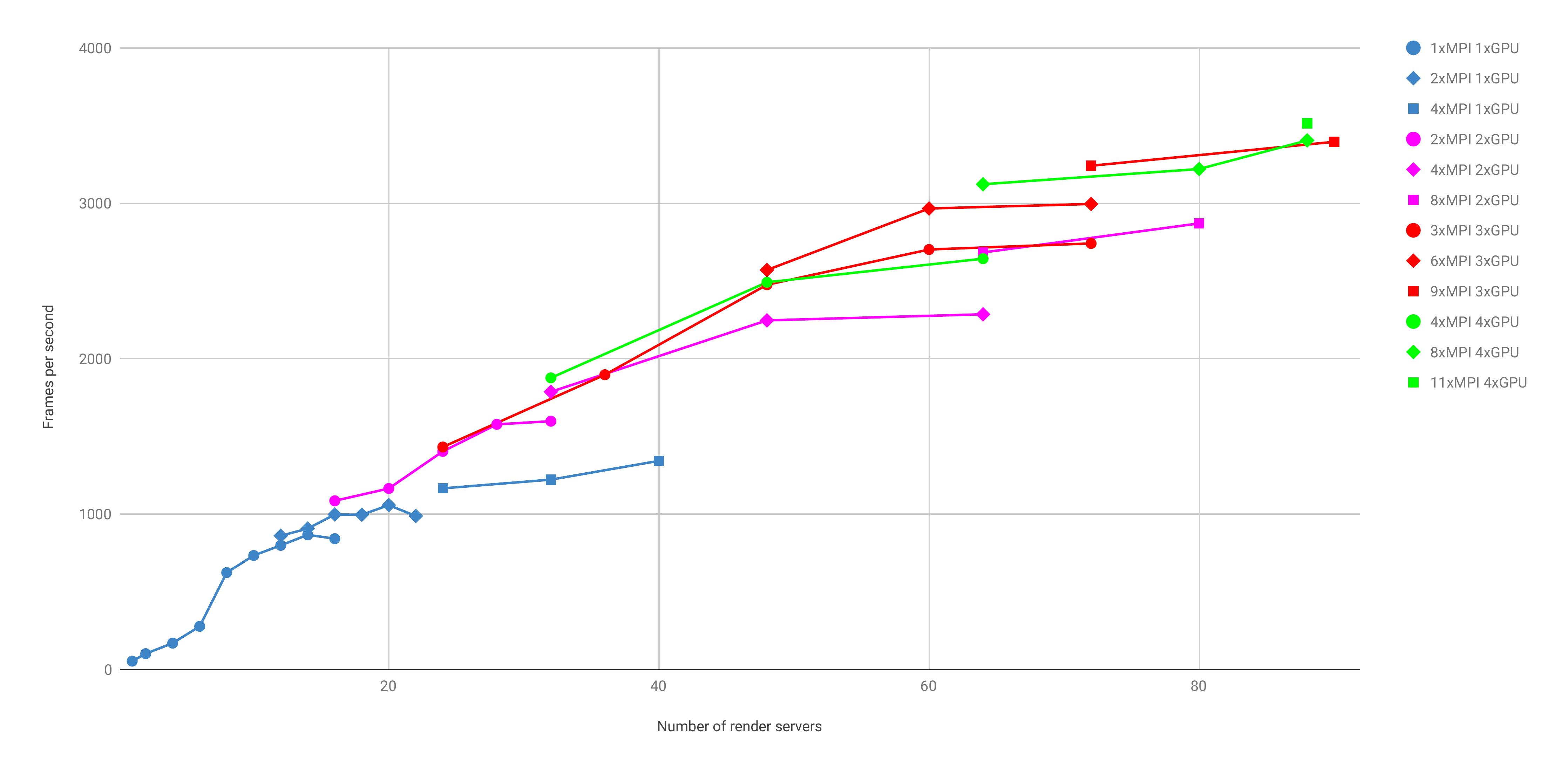} \vspace{0.3cm}
    \label{table:perf-mpi-gpu-worker}
\end{table}

\begin{table}[ht!]
    \centering
    \caption{
    GPU scaling - maximal throughput achieved with different number of V100 GPUs.}
    \vspace{0.3cm}
    \renewcommand{\arraystretch}{1.3}
    \begin{tabular}{@{}lllll@{}}
        \textbf{GPU count} & \textbf{1} & \textbf{2} & \textbf{3} & \textbf{4} \\ 
        \midrule
        \textbf{FPS} & $1342$ & $2870$ & $3395$ & $3514$ \\
    \end{tabular}
    \label{table:perf-gpu}
\end{table}

\begin{table}[ht!]
    \centering
    \caption{
    CPU scaling - maximal throughput achieved with different number of render servers (using 8 V100 GPUs).}
    \vspace{0.3cm}
    \renewcommand{\arraystretch}{1.3}
    \begin{tabular}{@{}lllllllllllll@{}}
        \textbf{Render servers} & \textbf{8} & \textbf{16} & \textbf{24} & \textbf{32} & \textbf{40} & \textbf{48} & \textbf{56} & \textbf{64} & \textbf{72} & \textbf{80} &
        \textbf{88} \\ 
        \midrule
        \textbf{FPS} & $736$ & $1215$ & $1457$ & $1876$ & $2290$ & $2628$ & $2847$ & $3065$ & $3284$ & $3370$ & $3438$ \\
    \end{tabular}
    \label{table:perf-worker}
\end{table}

\section*{Author contributions}

Maciek Chociej led the development of ORRB, designed and implemented the core systems, the rendering components and the client library.

Peter Welinder designed and implemented auxiliary channel rendering and was involved in shaping the renderer API.

Lilian Weng designed and implemented screen space tracking and auxiliary outputs.

\section*{Acknowledgements}

We used these open source libraries to build ORRB: GRPC, protobuf, MIConvexHull, pb\_Stl, UnityFBXExporter \& PostProcessingV2.

We would like to thank the following people at Unity Technologies: Danny Lange, Marwan Mattar \& Vilmantas Balasevicius for their help during development and debugging.

We would also like to thank the following people for providing feedback on earlier versions of this manuscript: Josh Tobin \& Wojciech Zaremba.

\medskip
{
\small
\bibliography{paper}
}

\end{document}